\documentclass[preprint,aps]{revtex4}
\usepackage{graphicx}  
                       
\newcommand{\bea}{\begin{eqnarray}}
\newcommand{\eea}{\end{eqnarray}}
\newcommand{\be}{\begin{equation}}
\newcommand{\ee}{\end{equation}}                                               
\newcommand{\Tr}{\mathrm{Tr}}
  
\begin{document}

\title{Goldstone boson currents in a kaon condensed CFL phase}

\author{A.~Gerhold and T.~Sch\"afer}
\affiliation{Department of Physics,
  North Carolina State University,
  Raleigh, NC 27695}
\author{A.~Kryjevski}
\affiliation{Nuclear Theory Center,
  Indiana University, 
  Bloomington, IN 47408}

\begin{abstract}
We study the stability of the kaon condensed color-flavor locked (CFL) 
phase of dense quark matter with regard to the formation of a non-zero
Goldstone boson current. In the kaon condensed phase there is an
electrically charged fermion which becomes gapless near $\mu_s^{(1)}
\simeq 1.35\Delta$ and a neutral fermion which becomes gapless near 
$\mu_s^{(2)}\simeq 1.61\Delta$. Here, $\mu_s=m_s^2/(2p_F)$ is the shift 
in the Fermi energy due to the strange quark mass $m_s$ and $\Delta$ 
is the gap in the chiral limit. The transition to the gapless phase 
is continuous at $\mu_s^{(1)}$ and first order at $\mu_s^{(2)}$. We 
find that the magnetic screening masses are real in the regime $\mu_s<
\mu_s^{(2)}$, but some screening masses are imaginary for $\mu_s>
\mu_s^{(2)}$. We show that there is a very weak current instability 
for $\mu_s>\mu_s^{(1)}$ and a more robust instability in a small window
near $\mu_s^{(2)}$. We show that in the Goldstone boson current phase all 
components of the magnetic screening mass are real. There is a range 
of values of $\mu_s$ below $2\Delta$ in which the magnetic gluon 
screening masses are imaginary but the phase is stable with respect 
to electrically neutral fluctuations of the gauge field.
\end{abstract}

\maketitle

\section{Introduction}
\label{sec_intro}

 Understanding the response of a paired fermion system to an external  
pair breaking field has been an area of intense theoretical and experimental 
research over the last couple of years. In dense quark matter the problem 
arises in connection with the effect of a non-zero strange quark mass, 
see \cite{Alford:2006fw} for recent reviews. If quarks are massless and 
the flavor symmetry is exact then the ground state of three flavor quark 
matter at high baryon density is the  color-flavor-locked (CFL) phase 
\cite{Alford:1999mk,Schafer:1999fe,Evans:1999at}. The CFL state involves
pairing between up and down, up and strange, as well as down and strange
quarks. In the real world the up and down quarks are approximately massless 
but the mass of the strange quark is comparable to the QCD scale parameter. 
At high baryon density the effect of the strange quark mass is governed by 
the shift $\mu_q = m_q^2/(2\mu)$ of the Fermi energy due to the mass. 

 A non-zero $\mu_s$ implies a non-zero pair breaking stress on $ud$ and
$ds$ pairs. Once CFL pairing is disrupted, an additional stress arises
due to the different electric charges of the $u$ and $d$ quark. There 
are two scales that govern the response to $\mu_s$, the mass $m_K$ of the 
lightest strange Goldstone boson, the kaon, and the gap $\Delta$ for 
fermionic excitations. When $\mu_s$ is equal to $m_K$ the CFL phase 
undergoes a transition to a phase in which the CFL order parameter 
rotates in flavor space and a kaon condensate appears \cite{Bedaque:2001je}.
If $\mu_s>\Delta$ gapless fermion modes appear in the spectrum and the CFL 
pairing pattern starts to get disrupted \cite{Alford:2003fq,Alford:2004hz}. 
We will refer to the phase with kaon condensation as the CFLK phase, 
and the gapless phases as the gCFL and gCFLK phase. 

 Gapless fermion modes in weakly coupled pair condensates cause instabilities 
in current-current correlation functions \cite{Wu:2003,Huang:2004bg} 
and these instabilities have been found in the gapless CFL phase
\cite{Casalbuoni:2004tb,Alford:2005qw,Fukushima:2005cm}. We have 
suggested that the instability is resolved by the formation of a 
non-zero Goldstone boson current 
\cite{Kryjevski:2005qq,Schafer:2005ym,Gerhold:2006dt}. In our previous
work we analyzed the Goldstone boson current in the kaon condensed
CFL phase, but we restricted the pairing ansatz to a flavor rotation
of the CFL state and did not properly implement charge neutrality 
in the Goldstone boson current state \cite{Kryjevski:2005qq,Schafer:2005ym}.
In \cite{Gerhold:2006dt} we allowed the CFL pairing pattern to be 
disrupted and constructed a charge neutral solution, but we did
not allow kaon condensation. 

 In the present work we study Goldstone boson currents in a kaon 
condensed CFL phase. We use a very general ansatz for the current, 
allow the CFL pairing pattern to be disrupted, and implement electric
and color charge neutrality. We also compute the magnetic screening 
masses. Our work remains restricted to states in which the Goldstone
boson current is small compared to the gap. This is expected to be 
the case as long as $\mu_s$ is close to the onset for gapless fermion
modes in the spectrum. The opposite limit of a current which is large 
compared to the gap in analyzed in studies of the 
Larkin-Ovchinnikov-Fulde-Ferrell (LOFF) state in three flavor QCD 
\cite{Alford:2000ze,Casalbuoni:2003wh,Casalbuoni:2005zp,Ciminale:2006sm,Mannarelli:2006fy}.

\section{Effective Lagrangian and dispersion laws}
\label{sec_eft}

 Our starting point is an effective lagrangian for gapped fermions 
interacting with gauge fields \cite{Kryjevski:2004jw,Kryjevski:2004kt}
\bea
\label{l_eff}
{\cal L } &=& 
     \Tr\left(\chi_L^\dagger(iv\cdot\partial -\hat{\mu}^L-A_e Q)\chi_L\right)
   + \Tr\left(\chi_R^\dagger(iv\cdot\partial -\hat{\mu}^R-A_e Q)\chi_R\right)
 \nonumber \\
 & & \mbox{}
  -i \Tr \left(\chi_L^\dagger \chi_L X v\cdot(\partial-iA^T)X^\dagger\right)
  -i \Tr \left(\chi_R^\dagger \chi_R Y v\cdot(\partial-iA^T)Y^\dagger\right)
 \nonumber \\
 & & \mbox{}
 -\frac{1}{2}\sum_{a,b,i,j,k} 
  \Delta_k\left(\chi_{L}^{ai}\chi_{L}^{bj}
                \epsilon_{kab}\epsilon_{kij}
               -\chi_{R}^{ai}\chi_{R}^{bj}
                \epsilon_{kab}\epsilon_{kij} + h.c. \right).
\eea
The notation agrees with our earlier work \cite{Gerhold:2006dt}.
$\chi_{L,R}^{ai}$ are left/right handed fermions with color 
index $a$ and flavor index $i$, $A_\mu$ are $SU(3)_C$ color gauge 
fields, and $\hat{\mu}^L=MM^\dagger/(2\mu)$, $\hat{\mu}^R=M^\dagger
M/(2\mu)$ are effective chemical potentials induced by the quark 
mass matrix $M$. The trace part of $\hat{\mu}^{L,R}$ corresponds
to a shift in the baryon chemical potential and will be neglected.
The matrix $Q=\mathrm{diag}({2\over3},-{1\over3},-{1\over3})$ is 
the quark charge matrix and $A_e$ is an electro-static potential. 
The fields $X,Y$ determine the flavor orientation of the left and 
right handed gap terms and transform as $X\to LXC^T$, $Y\to RYC^T$ 
under $(L,R)\in SU(3)_L \times SU(3)_R$ and $C\in SU(3)_C$, and 
$\Delta_k$ $(k=1,2,3)$ are the CFL gap parameters. From the lagrangian 
given in equ.~(\ref{l_eff}) we can read off the left handed Nambu-Gor'kov 
propagator,
\be
\label{prop}
  \left(
  \begin{array}{cc} 
     G_{(L)}^+   & \Xi_{(L)}^- \\ 
     \Xi_{(L)}^+ & G_{(L)}^- \end{array}
  \right)=\left(
  \begin{array}{cc}
    (p_0-p){\bf1}-\mathcal{X}_{(L),v}& \underline{\Delta} \\ 
    \underline{\Delta} & (p_0+p){\bf1}+ \mathcal{X}^T_{(L),-v}
  \end{array}
  \right)^{-1},
\ee
where $p=\vec v\cdot\vec p-\mu$, with the Fermi velocity $\vec v$.
The right handed propagator is obtained by replacing $(L)$ with $(R)$.
The components of the propagator are matrices in color-flavor 
space. We use a basis spanned by the Gell-Mann matrices $\lambda^A$
$(A=1,\ldots,8)$ and $\lambda_0=\sqrt{2\over3}{\bf 1}$. In this basis
\be
  \underline{\Delta}^{AB}=-\textstyle{1\over2}\Delta_{ab}\varepsilon_{ija}
  \varepsilon_{rsb}\lambda^A_{ir} \lambda^B_{js}, 
  \qquad \Delta_{ab}=\mathrm{diag}(\Delta_1,\Delta_2,\Delta_3)_{ab}.
\ee
The right and left handed gauge field vertices are (for $Y=X^\dag$)
\bea
  \mathcal{X}_{(L),v,AB}={\textstyle{1\over2}}
  \Tr\left[\lambda_A(\hat\mu^L+A_e Q)\lambda_B+
  i\lambda_A\lambda_B X\, v\cdot (\partial-iA^T) X^\dag \right],\\
  \mathcal{X}_{(R),v,AB}={\textstyle{1\over2}}
  \Tr\left[\lambda_A(\hat\mu^L+A_e Q)\lambda_B+
  i\lambda_A\lambda_B X^\dag\, v\cdot (\partial-iA^T) X \right].
\eea
Instead of using $\lambda_3$ and $\lambda_8$ it will turn out to be 
convenient to use the following linear combinations,
\be
  \lambda_I    = \lambda_3+{1\over\sqrt{3}}\lambda_8, \quad 
  \lambda_{II} = \lambda_3-\sqrt{3}\lambda_8.
\ee
We remark that $[\lambda_I,\lambda_7]=0$. We assume a maximal kaon 
condensate \cite{Bedaque:2001je},
\be
  X=\xi_{K^0}\equiv\exp\left({i\pi\over4}\lambda_6\right),  \label{xi}
\ee
and make the following ans\"atze for $A^{0T}$ and $\vec A^T$
\cite{Casalbuoni:1999wu,Kryjevski:2003cu},
\bea
  && A^{0T} = -{1\over2}\left[X^\dag(\hat\mu^L+A_e Q)X
     + X(\hat\mu^L+A_e Q)X^\dag\right]
     + \tilde A_I\lambda_I+\tilde A_7\lambda_7, 
\label{a0} \\
  && \vec A^T={\vec\jmath\over2}\left(c_I\lambda_I
  -{c_0\over\sqrt{6}}\lambda_0+c_7\lambda_7\right). 
\label{ai}
\eea
This particular flavor structure leads to diagonal matrices for $XA^T_\mu 
X^\dag$ and $X^\dag A^T_\mu X$. For $\tilde A_7=0$ the ansatz (\ref{ai}) 
for $\vec A^T$ is equivalent to setting
\be
  X=VU\xi_{K^0}U^\dag \label{xc}
\ee
with
\be
  U=\exp\left[-{i\over4} \vec\jmath\cdot\vec x \,c_7\lambda_{II}\right], 
   \quad 
  V=\exp\left[-{i\over2} \vec\jmath\cdot\vec x 
     \left(c_I\lambda_I-{c_0\over\sqrt{6}}\lambda_0\right)\right],
\ee
and taking the solution for $\vec A^T$ given by $\vec A^T={i\over2}
(X^\dag\vec\partial X+X\vec\partial X^\dag)$. For $\tilde A_7\neq0$ 
the computation of dispersion relations using the ansatz (\ref{xc}) 
is not straightforward because the $x$-dependence does not drop out. 
Therefore we will use the ansatz given by equs. (\ref{xi}) and (\ref{ai}) 
as our starting point. In the end we will see that the condition $\tilde 
A_7=0$ is indeed fulfilled, which means that the ansatz for the gauge
field (\ref{ai}) is equivalent to the ansatz for the Goldstone boson 
current (\ref{xc}). We remark that the ansatz of Ref.~\cite{Schafer:2005ym} 
corresponds to $c_0=c_I=0$, $c_7\neq0$, while the ansatz of 
Ref.~\cite{Kryjevski:2005qq} corresponds to $c_0$, $c_7\neq0$, $c_I=0$.

We find that the tadpole diagrams with an external $A_{1,2,4,5,6,II}$ 
vanish identically for any values of the parameters of our ansatz.
Therefore the corresponding neutrality conditions are automatically 
satisfied for the system under consideration.

We find the following dispersion laws in the left handed sector,
\bea
  &&\epsilon_{(L)1}={\mu_s\over2}-\tilde A_7+(2c_I+c_0)
      {\vec v\cdot\vec\jmath\over6}
  + \sqrt{\left(p+{2\tilde A_I\over3}-c_7
      {\vec v\cdot\vec\jmath\over2}\right)^2+\Delta_1^2}, \nonumber\\
  &&\epsilon_{(L)2}=-{\mu_s\over2}+\tilde A_7+(2c_I+c_0)
      {\vec v\cdot\vec\jmath\over6}
  +\sqrt{\left(p+{2\tilde A_I\over3}+c_7
      {\vec v\cdot\vec\jmath\over2}\right)^2+\Delta_1^2}, \nonumber\\ 
  &&\epsilon_{(L)3}={3\mu_s\over4}+\tilde A_I-{\tilde A_7\over2}
         -A_e-\left(2(c_I-c_0)+3c_7\right)
      {\vec v\cdot\vec\jmath\over12} \nonumber\\
  &&\qquad\quad+\ \sqrt{\left(p+{\mu_s\over4}-{\tilde A_I\over3}
         -{\tilde A_7\over2}
     + (2c_I-c_7){\vec v\cdot\vec\jmath\over4}\right)^2
     + \Delta_2^2},\nonumber\\
  &&\epsilon_{(L)4}=-{3\mu_s\over4}-\tilde A_I+{\tilde A_7\over2}
     +A_e-\left(2(c_I-c_0)+3c_7\right)
      {\vec v\cdot\vec\jmath\over12} \nonumber\\
  &&\qquad\quad+\ \sqrt{\left(p+{\mu_s\over4}-{\tilde A_I\over3}
     -{\tilde A_7\over2}
     -(2c_I-c_7){\vec v\cdot\vec\jmath\over4}\right)^2
     +\Delta_2^2},\nonumber\\
  &&\epsilon_{(L)5}={\mu_s\over4}+\tilde A_I+{\tilde A_7\over2} 
     -A_e-\left(2(c_I-c_0)-3c_7\right)
      {\vec v\cdot\vec\jmath\over12} \nonumber\\
  &&\qquad\quad+\ \sqrt{\left(p-{\mu_s\over4}-{\tilde A_I\over3}
     +{\tilde A_7\over2}
     +(2c_I+c_7){\vec v\cdot\vec\jmath\over4}\right)^2
     +\Delta_3^2},\nonumber\\
  &&\epsilon_{(L)6}=-{\mu_s\over4}-\tilde A_I-{\tilde A_7\over2}
     +A_e-\left(2(c_I-c_0)-3c_7\right)
     {\vec v\cdot\vec\jmath\over12} \nonumber\\
  &&\qquad\quad+\ \sqrt{\left(p-{\mu_s\over4}-{\tilde A_I\over3}
     +{\tilde A_7\over2}
     -(2c_I+c_7){\vec v\cdot\vec\jmath\over4}\right)^2
     +\Delta_3^2},\nonumber\\
  &&\epsilon_{(L)7,8,9}=\mathrm{complicated}.
\eea
The final three (``complicated'') dispersion laws cannot be expressed 
in simple form. They are electrically neutral and do not have much
impact on the dynamics, since they never become gapless for the values 
of $\mu_s$ which we will consider. The dispersion laws for the right 
handed sector can be obtained from the left handed ones by substituting 
$\tilde A_7\to-\tilde A_7$ and $c_7\to-c_7$. We note that for $c_I=c_0$ 
and $c_7=0$ the current does not shift the energy of the electrically 
charged modes, which is similar to the case without kaon condensate.

\section{Thermodynamic potential}
\label{sec_free}

\begin{figure}
\includegraphics[width=9cm]{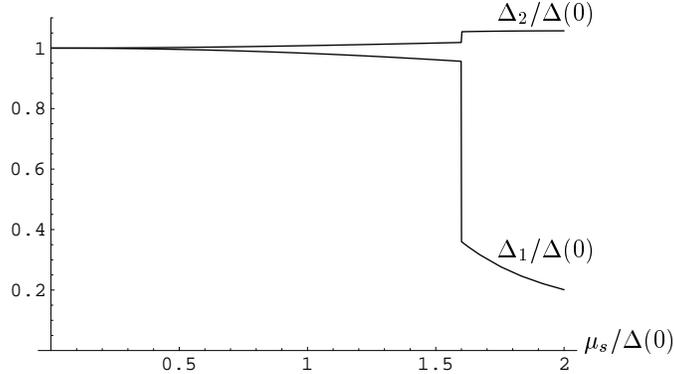}
\caption{Fermion gaps in the kaon condensed CFL phase as a function
of the effective chemical potential $\mu_s$. The figure shows $\Delta_2
/\Delta(0)$ (upper curve) and $\Delta_1/\Delta(0)$ (lower curve) as 
functions of $\mu_s/\Delta(0)$ for $\jmath=0$. Within our approximations
$\Delta_3=\Delta_2$. 
\label{figd1d2g}}
\end{figure}

 In order to evaluate the thermodynamic potential and study the 
phase structure we will make a number of approximations. In QCD 
gluons are dynamical fields, and gluon exchanges determine the 
value of the gap parameter. In the following we will treat the 
gauge fields and the gap as classical mean fields. The mean field
partition function reproduces low energy QCD fermion propagators 
as well as $n$-point functions of the currents and gauge fields
to leading order $O(\alpha_s^0)$ in the strong coupling constant 
and to $O(\mu_s^2)$ in the effective chemical potential. 

 In the mean field approximation the thermodynamic potential is 
given by
\be
\label{om1}
  \Omega={1\over G}(\Delta_1^2+\Delta_2^2+\Delta_3^2)
   +{(8c_I^2+c_0^2+6c_7^2)\mu^2 \jmath^2\over24\pi^2}-{\mu^2\over2\pi^2}
  \int dp\int {d\hat v\over 4\pi}
  \sum_{i=1}^9(|\epsilon_i|-|p|)-{A_e^4\over12\pi^2},
\ee
where the first term is the mean field potential for the gap field, the 
second is a tadpole term that arises from integrating out anti-fermions 
\cite{Son:1999cm}, and the third term is the one-loop effective potential 
of fermions coupled to background gauge and gap fields. We will keep only 
the leading order terms in $\Delta/\mu$ in the thermodynamic potential. 
In particular, we have written the integral over the fermion dispersion
relations in terms of the density of states, an integral $dp$ over the 
momentum transverse to the Fermi surface, and an angular integral 
$d\hat v$. We have included the $A_e^4$ term which is formally of higher 
order, but necessary to determine a unique solution to the electrical 
neutrality condition. 

 The integral over $p$ is ultraviolet divergent and requires a cutoff 
$\Lambda$. This divergence can be absorbed into the coupling constant 
$G(\Lambda)$. In the following $G$ refers to $G(\Lambda\!=\!\mu)$. The 
constant $G$ fixes the magnitude of the gap in the chiral limit, which 
we denote by $\Delta(0)$. We shall study the thermodynamic potential as 
a function of the dimensionless variables $\mu_s/\Delta(0)$ and $\jmath/
\Delta(0)$. 

 The integrals over $\epsilon_{1-6}$ are easily evaluated. The result is
\bea
  &&\Omega={1\over G}(\Delta_1^2+\Delta_2^2+\Delta_3^2)
      +{(8c_I^2+c_0^2+6c_7^2)\mu^2 \jmath^2\over24\pi^2}+R
      -{A_e^4\over12\pi^2}+\nonumber\\
  &&\qquad+\bigg[F\left(\Delta_1,{2\tilde A_I\over3},{\mu_s\over2}
      -\tilde A_7,{\jmath\, c_7\over2},(2c_I+c_0){\jmath\over6}\right)
   \nonumber\\
  &&\qquad\quad+\,F\bigg(\Delta_2,{\mu_s\over4}+{\tilde A_I\over3}
      +{\tilde A_7\over2},{3\mu_s\over4}+\tilde A_I-{\tilde A_7\over2}
      -A_e,(2c_I-c_7){\jmath\over4},\nonumber\\
  &&\qquad\qquad\qquad(2(c_I-c_0)+3c_7){\jmath\over12}\bigg)\nonumber\\
  &&\qquad\quad+\,F\bigg(\Delta_3,{\mu_s\over4}-{\tilde A_I\over3}
      +{\tilde A_7\over2},{\mu_s\over4}+\tilde A_I+{\tilde A_7\over2}
      -A_e,(2c_I+c_7){\jmath\over4},\nonumber\\
  &&\qquad\qquad\qquad(2(c_I-c_0)-3c_7){\jmath\over12}\bigg)
      +(\tilde A_7\leftrightarrow-\tilde A_7, c_7\leftrightarrow-c_7)\bigg],
\eea
where $R$ is the integral of the complicated modes,
\be
  R=-{\mu^2\over2\pi^2}\int dp\int {d\hat v\over 4\pi}
  \sum_{i=7}^9(|\epsilon_i|-|p|).
\ee
The function $F(\Delta,a_1,a_2,\jmath_1,\jmath_2)$ is defined by 
\bea
  &&F(\Delta,a_1,a_2,\jmath_1,\jmath_2)=-{\mu^2\over12\pi^2}
   \left[3\Delta^2\left(1+2\log\left(2\mu\over\Delta\right)\right)
    + 6a_1^2+2\jmath_1^2\right]  \nonumber\\
  &&\qquad+{\mu^2\over24\pi^2}\bigg\{\Theta(a_2-\jmath_2-\Delta)
      {1\over\jmath_2}
  \bigg[\lambda(\jmath_2)\left(\lambda(\jmath_2)^2+3\Delta^2\right) 
        \nonumber\\
  &&\qquad\quad+3\Delta^2(a_2-\jmath_2)
  \log\left({a_2-\jmath_2-\lambda(\jmath_2)\over a_2-\jmath_2
    +\lambda(\jmath_2)}\right)\bigg]
    +(\jmath_2\leftrightarrow-\jmath_2)\bigg\},  \label{ff} 
\eea
with $\lambda(\jmath_2)=\sqrt{(a_2-\jmath_2)^2-\Delta^2}$. Here, $a_1$
and $\jmath_1$ refer to the shift in the minimum of the dispersion
relation due to the gauge potentials and the current, and $a_2$
and $\jmath_2$ denote the shift in the energy. The equations for 
color and electrical neutrality and the equation $\partial\Omega/
\partial\Delta_2=\partial\Omega/\partial\Delta_3=0$ are solved by
\bea
  && \quad \tilde A_I=\tilde A_7=0, \qquad\Delta_3=\Delta_2,\nonumber\\
  && A_e=\max\left(0,-\Delta_2+{3\mu_s\over4}
     +\left(2|c_I-c_0|+3|c_7|\right){\jmath\over12}\right),
\eea
which greatly simplifies the expression for the thermodynamic potential.
The remaining equations for $\Delta_1$, $\Delta_2$ and $\jmath$ are 
solved by numerical minimization of the thermodynamic potential.

\begin{figure}
\includegraphics[width=9cm]{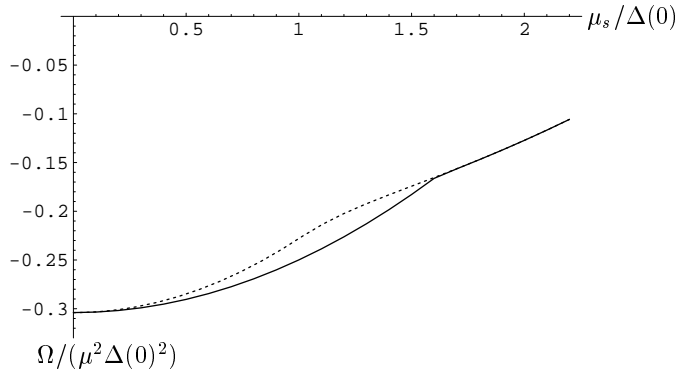}
\caption{Thermodynamic potential of the kaon condensed CFL phase $\Omega
/(\mu^2\Delta(0)^2)$ relative to the thermodynamic potential of neutral 
non-interacting quark matter  as a function of $\mu_s/\Delta(0)$ for 
$\jmath=0$. For comparison we also show the thermodynamic potential of 
the gCFL phase using the same approximations (dotted line). 
\label{figoms}}
\end{figure}

 First let us consider the case $\jmath=0$. Fig.~\ref{figd1d2g} shows 
the solution of the gap equations. We notice that even for small $\mu_s$ 
there is a small splitting between $\Delta_1$ and $\Delta_{2,3}$. The 
thermodynamic potential is shown in Fig.~\ref{figoms}, where we also 
compare to the result for the CFL-gCFL phase without a kaon condensate.
We find that $\epsilon_4$ becomes quadratic in $p$ for $\mu_s>\mu_s^{(1)}
=1.35\Delta(0)$, and that the mode $\epsilon_2$ becomes gapless for 
$\mu_s>\mu_s^{(2)}=1.607\Delta(0)$. In contrast to the CFL-gCFL phase 
transition, the phase transition at $\mu_s^{(2)}$ is first order. 
At $\mu_s=\mu_s^{(2)}$ there is a substantial drop in $\Delta_1$ 
and a small increase in $\Delta_{2,3}$. The thermodynamic potential 
is smooth at $\mu_s^{(1)}$ but has a kink at $\mu_s^{(2)}$. 

 We observe that near $\mu_s^{(2)}$ the thermodynamic potential of the 
gCFLK phase is very close to the thermodynamic potential of the gCFL phase 
($\Omega_{gCFLK}<\Omega_{gCFL}$ by a very small amount). It is not entirely 
clear why this is the case. Near $\mu_s^{(2)}$ the gauge potentials become 
quite large and our approximations are breaking down. Also, since some of 
the gaps become quite small it is not clear whether the correct CFLK state 
is a simple flavor rotation of the CFL state. We note that near $\mu_s^{(2)}$ 
the thermodynamic potential in the gCFL phase plotted in Fig.~\ref{figd1d2g} 
is lower than the result of Alford et al.~\cite{Alford:2004hz}. This 
difference is also due to higher order terms. In the gCFL phase the 
thermodynamic potential in equ.~(\ref{om1}) differs from the functional 
used by Alford et al.~by terms of $O(\mu_s^4)$. 

\begin{figure}
\includegraphics[width=9cm]{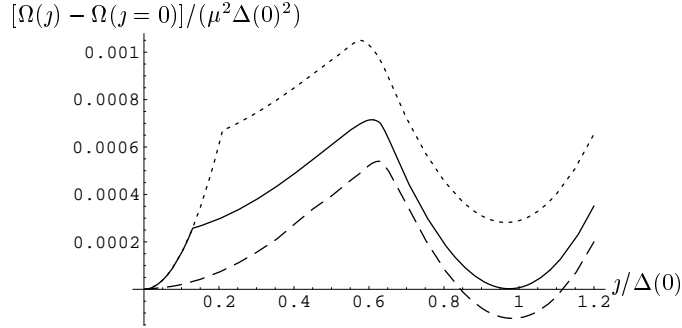}
\caption{Thermodynamic potential as a function of the Goldstone
boson current near $\mu_s^{(2)}$. The figure shows $[\Omega(\jmath)
-\Omega(\jmath=0)]/(\mu^2\Delta(0)^2)$ at $\mu_s=1.600\Delta(0)$ 
(dotted), $\mu_s=1.605\Delta(0)$ (continuous) and $\mu_s=1.610
\Delta(0)$ (dashed). \label{figomj}}
\end{figure}

  Next we study the effect of a Goldstone boson current. We first 
consider the region $\mu_s \simeq \mu_s^{(1)}$, which was also 
investigated in \cite{Schafer:2005ym,Kryjevski:2005qq}. If we take 
into account the condition of electrical neutrality we find that the 
expectation value of the current becomes extremely small. The competition 
between the $A_e^4$  and $\mu^2\jmath^2$ terms leads to a minimum of 
$\Omega$ at $\jmath\sim({3\over4}\mu_s-\Delta_2)^3/\mu^2$, corresponding 
to a contribution to $\Omega$ of the order $({3\over4}\mu_s-\Delta_2)^6
/\mu^2$. A detailed calculation of the current is presented in the 
appendix. Since the current is so small, we shall neglect its effect
in the remainder of this section. 

  Now consider the region $\mu_s\simeq \mu_s^{(2)}$. Here the optimal 
value for $c_7$ turns out to be zero, because $c_7$ does not contribute 
to the shift of the energy of $\epsilon_{2}$. If we set $c_I=c_0$ the 
current shifts only the energy of the electrically neutral mode. This 
should be the preferred current if electrical neutrality is enforced.
We set $c_I=1.2$ and $c_0=1$, which leads to an even stronger minimum 
of the thermodynamic potential at finite $\jmath$.

  Fig.~\ref{figomj} shows the thermodynamic potential as a function of 
$\jmath$ for different values of $\mu_s$ in the vicinity of $\mu_s=1.6
\Delta(0)$. The kinks come from a discontinuity in $\Delta_{1,2}(\jmath)$, 
similar to the discontinuity in $\Delta_{1,2}(\mu_s)$ shown in 
Fig.~\ref{figd1d2g}. We observe that the thermodynamic potential develops 
a nontrivial minimum if $\mu_s>1.605\Delta(0)$, which is slightly below 
the value of $\mu_s$ where the mode $\epsilon_2$ becomes gapless at zero 
current. We remark that the window where the thermodynamic potential has 
a nontrivial minimum is rather small. For $\mu_s$ slightly above $1.615
\Delta(0)$ we find that the global minimum is again at $\jmath=0$. We 
note, however, that the some of the gaps are very small, the current at 
the non-trivial minimum is rather large, and our approximations are not 
reliable. Since the currents are large the correct ground state may well 
be of the LOFF type and involve multiple currents and complicated crystal 
structures as in \cite{Rajagopal:2006ig}. In addition to that, 
inhomogeneities of the amplitude of the order parameter may play a 
role \cite{Iida:2006df,Giannakis:2006gg}.

\section{Meissner masses}
\label{sec_mei}

\begin{figure}
\includegraphics[width=9cm]{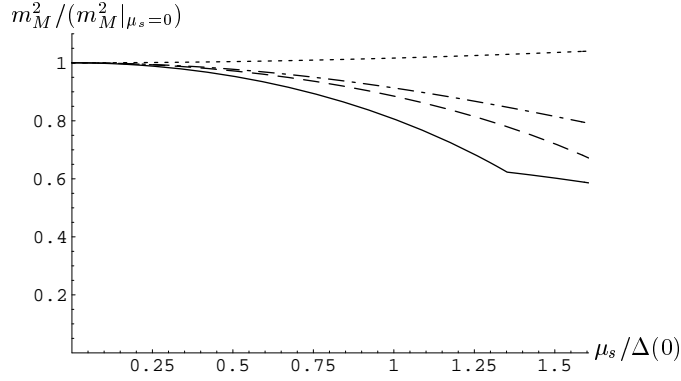}
\caption{Magnetic gluon screening (Meissner) masses in the kaon condensed
CFL phase as a function of the effective chemical potential $\mu_s$. The
figure shows squared Meissner masses $m_{M11}^2=m_{M22}^2=m_{M44}^2=m_{M55}^2$ 
(continuous), $m_{M66}^2=m_{M(1)}^2$ (dashed), $m_{M77}^2$ (dotted), 
$m_{M(2)}^2$ (dash-dotted) in units of $m_M^2|_{\mu_s=0}\equiv 
(21-8\log2)g^2\mu^2/(108\pi^2)$ as functions of $\mu_s/\Delta(0)$. 
\label{figm0}}
\end{figure}

 In this section we shall study the stability of the CFLK and gCFLK
phase with and without a Goldstone boson current with regard to 
chromomagnetic instabilities. At leading order in $\alpha_s$ the 
screening masses are given by a one-loop diagram involving 
Nambu-Gor'kov propagators. We have
\bea
(m^2_M)_{ab}^{ij} &=& {g^2\mu^2\over2\pi^2} \delta_{ab}\delta^{ij}
   + {g^2\mu^2 \over16\pi^2} \lim_{\vec{k}\to 0}\, \lim_{k_0^{}\to0}
   \int dp\int {d\hat v\over4\pi}\int {dp_0\over{2\pi}}
       \nonumber \\[0.2cm]
  & &   \hspace{0.0cm}\mbox{} \times\mathrm{Tr}
      \Big[ G^+(p) V_{(L)a}^i G^+(p+k)V_{(L)b}^j
         +  G^-(p)\tilde V_{(L)a}^i G^-(p+k)\tilde V_{(L)b}^j \nonumber \\
  & & \hspace{0.5cm}\mbox{}
         + \Xi^+(p) V_{(L)a}^i\Xi^-(p+k)\tilde V_{(L)b}^j
         + \Xi^-(p)\tilde V_{(L)a}^i \Xi^+(p+k) V_{(L)b}^j
         + (R\leftrightarrow L)\Big],
\eea
where we have defined the vertices
\bea
  (V_{(L)a}^i)_{AB}=\textstyle{1\over2}
    \Tr[\lambda_A\lambda_B X\lambda_a^TX^\dag]v^i,
    &\qquad& 
  (\tilde V_{(L)a}^i)_{AB}=-\textstyle{1\over2}
    \Tr[\lambda_B\lambda_AX\lambda_a^TX^\dag] \tilde v^i,  \nonumber\\
  (V_{(R)a}^i)_{AB}=\textstyle{1\over2}
    \Tr[\lambda_A\lambda_BX^\dag\lambda_a^TX]v^i,
    &\qquad& 
   (\tilde V_{(R)a}^i)_{AB}=-\textstyle{1\over2}
    \Tr[\lambda_B\lambda_AX^\dag\lambda_a^TX] \tilde v^i,\nonumber
\eea
with $\tilde v^i=-v^i$. We remark that right handed and left handed 
quark propagators are identical because of $\tilde A_7=c_7=0$.

 We find that the mass matrix  in the 3-8-sector is not diagonal, the 
mixing angle being equal to $\pi/6$. We denote the corresponding 
eigenvalues by $m^2_{M(1)}$ and $m^2_{M(2)}$. We find that some 
of the Meissner masses are equal,
\be
   m_{M11}^2=m_{M22}^2=m_{M44}^2=m_{M55}^2, \quad m_{M66}^2=m_{M(1)}^2.
\ee
Fig.~\ref{figm0} shows the Meissner masses as functions of $\mu_s$ for 
$\mu_s<\mu_s^{(2)}=1.607\Delta(0)$. The kink in $m_{M11}$ at $\mu_s
=\mu_s^{(1)}$ results from the kink in $A_e$. We observe that the 
Meissner masses in the gapless CFLK phase are real, even in the 
absence of a current.

\begin{table}
\begin {tabular}{|c||c|c|c|c|}\hline
 & $m_{M11}^2$  & $m_{M66}^2$ & $m_{M77}^2$ & $m_{M(2)}^2$ \\ \hline\hline
transv. & 0.16 & 0.59  &  1.09  &  0.21  \\ \hline
long.   & 0.31 & 0.87  &  1.05  &  0.46  \\ \hline
\end{tabular}
\caption{Meissner masses at the minimum of $\Omega(\jmath)$ for $\mu_s=
 1.605\Delta(0)$, in units of $m_M^2|_{\mu_s=0}$.
\label{tabm}}
\end{table}

 For $\mu_s>\mu_s^{(2)}$ and zero current the Meissner masses 
$m_{M11}^2=m_{M22}^2=m_{M44}^2=m_{M55}^2$ as well as $m_{M(1)}^2$
receive a negative contributions proportional to $-1/\sqrt{\mu_s^2-4
\Delta_1^2}$. This is the usual chromomagnetic instability for zero 
current. In the presence of a finite current $\vec \jmath$ we may decompose 
the Meissner masses into a longitudinal and a transverse component,
\be
  (m_M^2)^{ij}=m_{M\perp}^2(\delta^{ij}-\hat \jmath^i\hat \jmath^j)
   +m_{M\parallel}^2\hat \jmath^i\hat \jmath^j.
\ee
The values of the Meissner masses at finite current just above the phase 
transition at $\mu_s=1.605\Delta(0)$ are shown in Table \ref{tabm}. We 
note that all these values are positive. We note that for $\mu_s>1.615
\Delta(0)$ there is no Goldstone boson current and the magnetic 
screening masses in the CFLK phase are tachyonic. This is an unusual
situation because stability with respect to a Goldstone boson current
implies that the phase is stable with respect to electrically neutral 
fluctuations of the gauge field. We will comment on this regime 
in the conclusions. 

\section{Conclusions}
\label{sec_sum}

 We have studied the stability of a kaon condensed CFL phase with 
respect to the formation of a Goldstone boson current. We have 
computed the thermodynamic potential and the gluon screening masses
as a function of the effective chemical potential $\mu_s = m_s^2
/(2p_F)$. Our starting point is an effective lagrangian of fermions 
coupled to a mean-field gap term and background gauge potentials. 
Our methods are reliable provided the gauge potentials and currents
are smaller than the gaps. There are three important scales that appear 
in the problem, $\mu_s=m_K$ is the onset of kaon condensation, $\mu_s=
\mu_s^{(1)}=1.35\Delta(0)$ is the point where a charged fermion mode 
becomes (almost) gapless, and $\mu_s=\mu_s^{(2)}=1.61\Delta(0)$ is the 
point where a neutral fermion mode becomes gapless. The CFLK-gCFLK 
transition at $\mu_s^{(1)}$ is smooth while the transition at 
$\mu_s^{(2)}$ is first order. 

 We find that the kaon condensed CFL phase is more stable with 
respect to chromo-magnetic instabilities than the CFL phase 
\cite{Zhang:2006bf,Warringa:2006dk}. The gluon screening masses 
are real in the regime $\mu_s<\mu_s^{(2)}$, despite the presence 
of an almost gapless mode. There is a weak Goldstone boson current 
instability for $\mu_s>\mu_s^{(1)}$. The magnitude of the current 
is suppressed by the constraint of electric charge neutrality. 

 We find that the gluon screening masses become imaginary 
in the CFLK phase for $\mu_s>\mu_s^{(2)}$. There is a Goldstone 
boson current instability that develops at this point, and 
the screening masses in the presence of a Goldstone boson current 
are positive. However, the Goldstone boson current phase only 
exists in a very small window above $\mu_s^{(2)}$. Beyond that 
point the CFLK phase is stable with respect to the formation 
of Goldstone boson currents, but the gluon screening masses
are imaginary. This is somewhat puzzling since the Goldstone
boson currents are equivalent to non-zero gauge fields, see
equ.~(\ref{ai}). The difference between the second derivative 
of the thermodynamic potential with respect to the Goldstone
boson currents and the screening masses is that the former 
are computed at constant (zero) electric charge, while the 
latter are computed at constant electro-static potential. In 
this sense stability with respect to Goldstone boson currents 
is the physically relevant criterion. The fact that the 
Meissner masses are imaginary may nevertheless imply that 
other instabilities are present. In this regime some of the 
gaps are small compared the background fields and it may 
be necessary to consider states with multiple currents, such 
as the crystalline LOFF state 
\cite{Mannarelli:2006fy,Casalbuoni:2005zp,Rajagopal:2006ig}.

Acknowledgments: This work is supported in part by the US Department
of Energy grant DE-FG-88ER40388 (T.S.~and A.G.) and DE-FG02-87ER40365 
as well as the National Science Foundation grant PHY-0244822 (A.K.).
We would like to thank J.~Kapusta for pointing out a sign mistake 
in the second part of equ.~(23) in \cite{Kryjevski:2004jw}.

\appendix
\section{Charge neutrality and Goldstone boson current for $\mu_s
>\mu_s^{(1)}$}
\label{sec_app}

 In this appendix we study in more detail the current instability in the 
regime $\mu_s>\mu_s^{(1)}\simeq 4\Delta/3$. For simplicity we only consider 
a pure hypercharge current with $V=1$. Including the baryon current 
will only change some numerical coefficients. Our analysis extends
the analysis of \cite{Kryjevski:2005qq,Schafer:2005ym} to properly 
implement charge neutrality in the Goldstone boson current state.

The instability is driven by the lowest fermion mode in the spectrum. 
The dispersion relation is given by
\be
\label{disp_ax}
\omega_p = \Delta +\frac{(p-l_0)^2}{2\Delta}-\frac{3}{4}
  \mu_s +A_e -\frac{1}{4}\vec{v}\cdot\vec{\jmath}_K,
\ee
where  $A_e$ is the electron chemical potential, $\mu_s=m_s^2/(2p_F)$ 
and $l_0=(\mu_s-\vec{v}\cdot\vec{\jmath}_K)/4$. The contribution of 
gapless modes to the thermodynamic potential is
\be 
 \Omega =-2 \frac{\mu^2}{2\pi^2}\int dp \int 
   \frac{d\hat{v}}{4\pi} \;\omega_p \theta(-\omega_p) ,
\ee
where the factor 2 is a degeneracy factor and $d\hat{v}$ is an 
integral over the Fermi surface. The charge density of gapless
modes is 
\be 
 q = 2 \frac{\mu^2}{2\pi^2}\int dp \int 
 \frac{d\hat{v}}{4\pi} \;\theta(-\omega_p) .
\ee
The electron contribution to the thermodynamic potential and the 
charge density is  
\be 
\Omega_e =-\frac{A_e^4}{12\pi^2}, \hspace{1.5cm} 
 q_e     =-\frac{A_e^3}{3\pi^2}.
\ee
We introduce dimensionless variables
\be 
 x = \frac{\jmath_K}{a\Delta}, \hspace{1cm}
 h = \frac{3\mu_s-4\Delta}{a\Delta},\hspace{1cm}
 h_e = \frac{4A_e}{a\Delta},
\ee
where $a$ is a numerical coefficient that depends on the 
parameters of the effective theory and the number of currents
that are turned on. For a pure kaon current 
\be
 a = \frac{2}{15^2 c_\pi^2 v_\pi^4}, 
\ee
where $c_\pi=(21-8\log(2))/36$ is the numerical constant that 
appears in the weak coupling result for the pion decay constant
$f_\pi$ and $v_\pi^2=1/3$ is the square of the Goldstone boson velocity. 
The thermodynamic potential and charge density can be written as 
\bea 
\Omega (h,h_e,x) &=& \frac{\mu^2\Delta^2}{\pi^2}
            \left( Cf_{h-h_e}(x)- C_e h_e^4 \right),\\
 q(h,h_e,x)      &=& \frac{\mu^2\Delta}{\pi^2}
            \left( Kg_{h-h_e}(x)- K_e h_e^3 \right) 
\eea
with 
\bea 
 f_h(x) &=& x^2-\frac{1}{x}\left[
   (h+x)^{5/2}\Theta(h+x) - (h-x)^{5/2}\Theta(h-x) \right],\\
 g_h(x) &=&  \frac{1}{x}\left[
   (h+x)^{3/2}\Theta(h+x) - (h-x)^{3/2}\Theta(h-x) \right].
\eea
The numerical coefficients are given by
\be
 C = \frac{2}{15^4c_\pi^3 v_\pi^6}
  \frac{\mu^2\Delta^2}{\pi^2},\hspace{1.5cm}
 K = \frac{10C}{a}
\ee
and 
\be
C_e = \frac{1}{3\cdot 4^5}\left(\frac{\Delta^2}{\mu^2}\right),\hspace{1.5cm}
K_e = \frac{16C_e}{a}.
\ee
We first solve the neutrality condition in order to find $h_e=h_e(h,x)$. 
We have $h_e=0$ if $h+x<0$. For $h+x>0$ we can use $K_e\ll 1$ in order 
to write 
\be
 h_e = h+x +\delta h_e
\ee
with $\delta h_e\ll h_e$. We get 
\be 
\delta h_e = \left(\frac{K_e}{K}\right)^{2/3} x^{2/3}(h+x)^2.
\ee
Inserting this result into the energy density gives 
\be 
\Omega = \frac{\mu^2\Delta^2}{\pi^2} \left\{
 C \left[ x^2 -\frac{1}{x} \delta h_e^{5/2}\Theta(\delta h_e)\right] 
  - C_e (h+x)^4 + \ldots \right\}.
\ee
The contribution from gapless fermions is proportional to 
$x^{2/3}$ for small $x$ so the instability is still present. 
If $K_e,C_e\ll 1$ the minimum of the thermodynamic potential is 
determined by the balance between the electron term
and the current contribution. We get
\be 
 x = \frac{2C_e}{C} h^3
   \sim \left(\frac{\Delta^2}{\mu^2}\right) h^3 .
\ee
This implies that a Goldstone current is formed, but the 
magnitude of the current is suppressed as compared to the 
result given in \cite{Kryjevski:2005qq,Schafer:2005ym}.


\end{document}